\documentclass[final,3p,times,twocolumn]{elsarticle}

 \usepackage{graphics}

\usepackage{amssymb}

\usepackage{lineno}

\usepackage{subfigure}

\journal{Nuclear Instruments and Methods A }

\begin{document}

\begin{frontmatter}




\title{The Online Calibration, Operation, and Performance of the CMS Pixel Detector}
\author[a1]{B.~Kreis\corref{cor1}}
\ead{kreis@cern.ch}
\author{for the CMS Collaboration}
\cortext[cor1]{Corresponding author.}
\address[a1]{Newman Laboratory for Elementary Particle Physics, Cornell University, Ithaca, NY 14853, USA}

\begin{abstract}
The CMS pixel detector consists of approximately 66 million silicon pixels whose analog signals are read out by 15,840 programmable Readout Chips. With the recent startup of the LHC, the detector is now collecting data used for precise vertexing and track-finding. In preparation for data taking, the detector's Readout Chips and their supporting readout and control electronics were calibrated.  The calibration that has taken place since the detector's installation in the summer of 2008 will be described.  These calibrations focused on the optimization of the Readout Chips' thresholds and analog response.  The operation of the detector during the early running of the LHC will also be discussed.  The calibrations that are performed on a regular basis and a mechanism to handle the readout of large beam background events will be described.
\end{abstract}


\end{frontmatter}


\section{Introduction}
\label{sec:1}

The Compact Muon Solenoid (CMS) pixel detector is the innermost tracking device of the CMS experiment located at the Large Hadron Collider (LHC) at CERN.  The detector consists of approximately 66 million ``n on n'' silicon pixels that are arranged in three barrel layers (BPix) and four forward disks (FPix) to provide coverage over the pseudorapidity range $|\eta|<2.5$.  The detector reads out analog pulse heights so that signal interpolation across pixels can be used to achieve the hit position resolution required for CMS vertexing and track-finding \cite{CMS, ttdr}.  

The pixel detector was installed and commissioned in 2008 \cite{danekComm}.  The detector was calibrated in the time leading up to and throughout the commissioning in 2008, and its performance was studied with cosmic ray muons \cite{cosmic}.  Further calibrations were performed in 2009 as the LHC prepared to deliver collisions.  These calibrations, which are described in Sections \ref{sec:3} and \ref{sec:4}, mainly addressed the thresholds and analog response of the detector.  The detector has been operating since collisions began in December 2009.  Parts of the data acquisition system are recalibrated on a regular basis to account for environmental changes and to monitor the detector's status.  These calibrations are described in Section \ref{sec:5}. Readout problems due to beam background were discovered as soon as collisions were delivered.  An update to the firmware of one device in the data acquisition system addressed this problem.  This is described in Section \ref{sec:6}.

\section{Pixel Data Acquisition}
\label{sec:2}
Groups of 4,160 pixels, making up 80 rows and 26 pairs of columns, known as double-columns, are read out by the PSI46 Readout Chip (ROC) \cite{ROC}.  After amplification and shaping, zero-suppression is performed on the ROC with a comparator for each pixel.  When a signal crosses the comparator's threshold, it is considered as a hit, and the analog pulse height, the address of the pixel, and the bunch crossing number are stored in buffers dedicated to its double-column for the latency time of the CMS first level trigger.  The ROC reads out a single, 25 nanosecond wide bunch crossing; hits are validated by the trigger and sent on to the pixel data acquisition system if the bunch crossing number of the hit and the trigger match.  

The ROC has 21 8-bit digital-to-analog converters (DACs), five 4-bit DACs and one 3-bit DAC that influence various aspects of the readout.  In addition, each pixel has four bits, called trim bits, that influence the comparator's threshold, and one bit to mask the pixel if needed.   The DAC settings are programmed before running the detector.  Calibration signals with an amplitude set by an 8-bit DAC known as VCal can be injected through a capacitor connected to the amplifier input node.  On average, one VCal DAC unit corresponds to 65.5 electrons with an offset of -414 electrons \cite{danekComm}.  This conversion is used throughout.  

The ROCs are read out and controlled by VME devices in the CMS electronics room.  The analog electrical signals from groups of ROCs are converted to an optical signal and sent to the Front End Driver (FED) VME device \cite{FEDpaper}.  The FED digitizes the signals and builds event fragments from the hits corresponding to one first level trigger.

\section{Threshold Calibrations}
\label{sec:3}

The thresholds of the detector are important performance parameters because they influence the cluster size, and therefore, the hit position resolution.  The threshold of a pixel's comparator depends on its four trim bits and two 8-bit DACs on its ROC known as VcThr and Vtrim.  The impact of these settings on the threshold of pixel $i$'s comparator, $\mathrm{Thr}_i$, is roughly given by,
\begin{equation}
\label{eq:trim}
\mathrm{Thr}_i  = C_{0} - C_{1} \mathrm{VcThr} - C_{2} \mathrm{Vtrim} \left(15-\mathrm{trim bits}_i\right),
\end{equation}
where $C_{0}$, $C_{1}$, and $C_{2}$ are positive constants.  As seen in Eq. \ref{eq:trim}, VcThr applies an offset to the threshold of every pixel on the ROC, and Vtrim determines how much influence the trim bits on the ROC have.

Due to time-walk, the smallest signals that cross threshold may do so in a bunch crossing following the triggered one in which the charge was actually deposited \cite{ROC}.  For this reason,  two thresholds are defined for each pixel.  The first is the absolute threshold, which is the charge required to cross threshold independent of the time at which it does so. It is precisely equal to the comparator's threshold.  The second is the in-time threshold, which is the charge required to cross threshold in the same bunch crossing as the one in which the charge was deposited.  The absolute threshold is relevant when studying occupancy, noise, and cross-talk, and the in-time threshold is relevant when studying hit reconstruction.

\begin{figure}[ht] 
  \begin{center}
	 \resizebox{7.5cm}{!}{\includegraphics{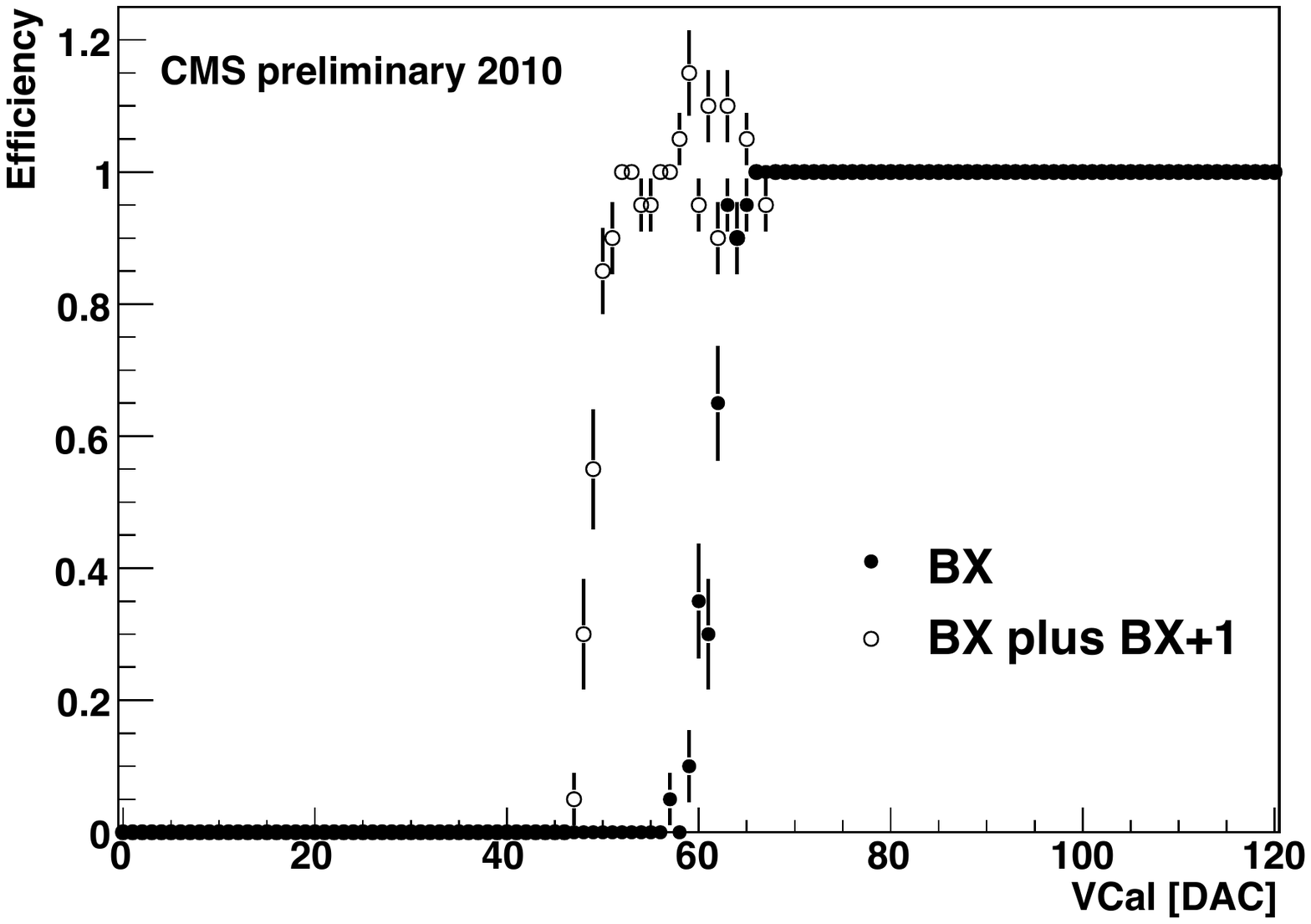}}
  \caption{S-Curve from the in-time bunch crossing (BX) (full circles) and sum of S-Curves from the in-time bunch crossing and the following one (open circles). The latter curve can exceed $100\%$ efficiency because it is the sum of two efficiency curves; this has a negligible effect on the fit.}
  \label{fig:SCurveInAbs}
  \end{center}
\end{figure}

The in-time thresholds depend on the timing of the detector's clock with respect to the LHC's collisions, however they can be estimated using charge injection.  The timing of the charge injection is set so that the maximum injected charge crosses threshold approximately five nanoseconds into the bunch crossing.  Both the absolute threshold and approximate in-time threshold of a pixel are measured using so-called S-Curves.  An S-Curve is the efficiency for injected charge to cross threshold in a specified bunch crossing versus the injected charge.  The in-time threshold is taken as the location of the turn-on of the S-Curve from the bunch crossing in which the charge was injected, that is, the in-time bunch crossing (see Fig. \ref{fig:SCurveInAbs}).  The absolute threshold is taken as the location of the turn-on of the sum of S-Curves from the in-time bunch crossing and the following one (see Fig. \ref{fig:SCurveInAbs}).    The location of a turn-on is taken as the injected charge at which an error function fit to the (summed) S-Curve(s) reaches $50\%$ efficiency.

\subsection{Threshold Trimming}
\label{subsec:31}

In the first step of the threshold calibration, either the in-time or absolute thresholds on each ROC were adjusted to the same VCal value by tuning VcThr, Vtrim, and the trim bits.  This procedure is known as ``trimming''.  The BPix absolute thresholds were trimmed using an algorithm described elsewhere \cite{BPixTrim}.  The FPix in-time thresholds were trimmed using an alternative algorithm.  

In the first step of the FPix trimming algorithm, the settings for VcThr and Vtrim are determined with an iterative algorithm that can be applied to a small but representative subset of the pixels on each ROC ($\sim2\%$ distributed across the ROC) to save time.  In each iteration, changes in the thresholds resulting from small changes in VcThr, Vtrim, and the trim bits are measured using S-Curves.   The next values for VcThr, Vtrim, and the trim bits are then solved for using a first order Taylor expansion of Eq. \ref{eq:trim} for each pixel and some additional requirements to constrain the problem.  Typically, four iterations are required to obtain satisfactory settings for VcThr and Vtrim.  In the second and final step of the FPix trimming algorithm, the threshold of every pixel is measured, and then based on the average influence of the subset of trim bits measured in the previous step, the final trim bit value of every pixel is chosen.

\begin{figure}[ht] 
  \begin{center}
	 \resizebox{7.5cm}{!}{\includegraphics{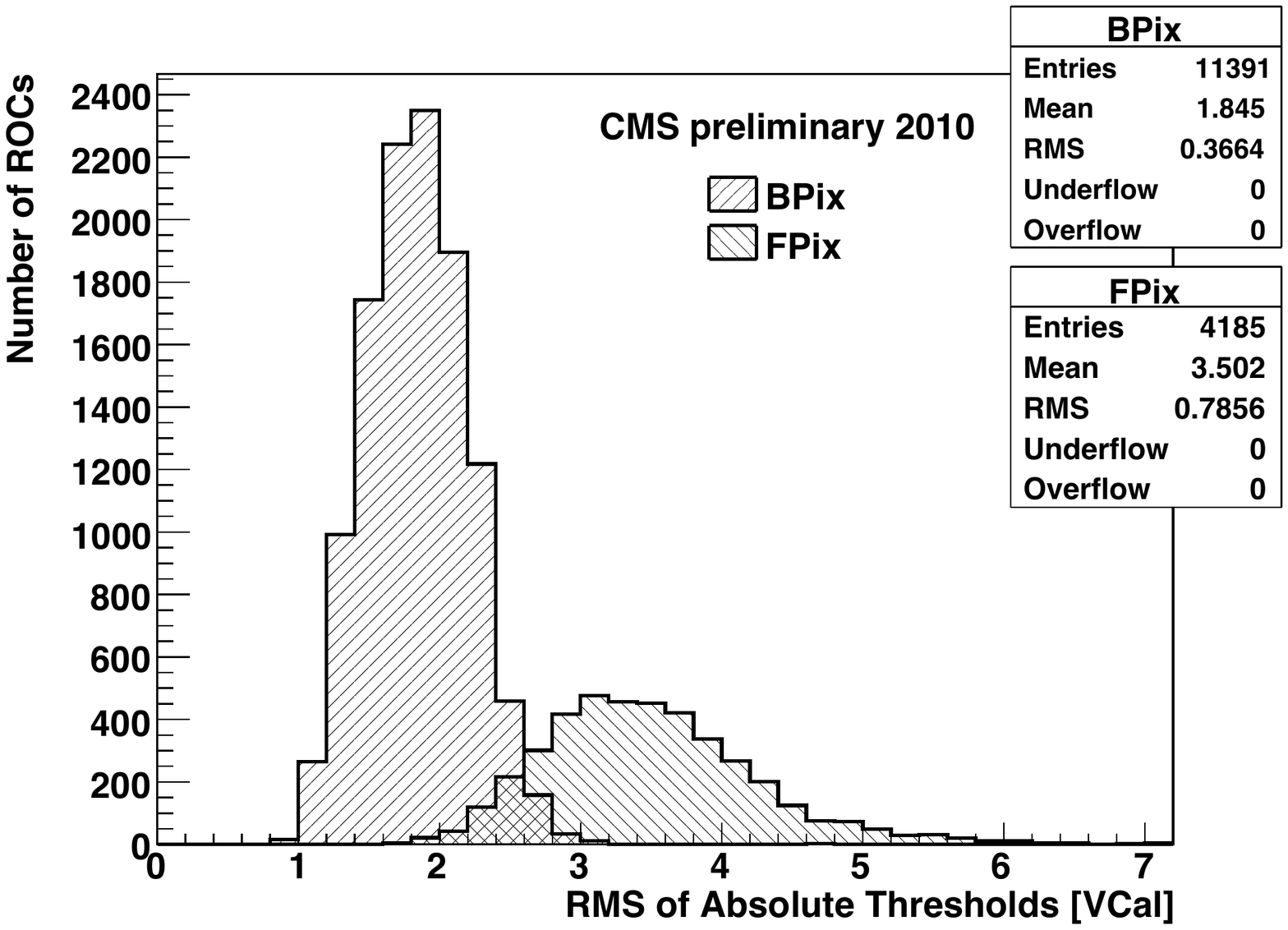}}
  \caption{Histogram of the RMS of the absolute threshold on each ROC.  The RMS is computed from $\sim2\%$ of the pixels on the ROC.}
  \label{fig:RMS}
  \end{center}
\end{figure}

A histogram of the ROC absolute threshold RMS is shown in Fig. \ref{fig:RMS}.  The RMSs are several times smaller than the variation in VCal \cite{danekComm}.  The FPix RMSs are slightly larger than those of the BPix because the in-time thresholds, rather than the absolute thresholds, were trimmed and because the calibration described in Section \ref{subsec:32} was performed after trimming.

\subsection{In-time Threshold Calibration}
\label{subsec:32}

The in-time thresholds depend on the amount of time-walk introduced in the amplification and shaping that occurs before the signal reaches the comparator.  This depends on Vana, an 8-bit DAC that regulates the voltage applied to the analog part of the ROC.  Vana was set in such a way that balanced the desire to minimize time-walk with the need to keep the current drawn by analog part of ROC, or ``analog current'', at a reasonable level.  The Vana setting for each BPix ROC was determined during module testing by directly measuring the analog current drawn by the ROC as a function of Vana and then choosing the Vana that corresponded to 24 mA. 

\begin{figure}[ht] 
  \begin{center}
	  \resizebox{7.5cm}{!}{\includegraphics{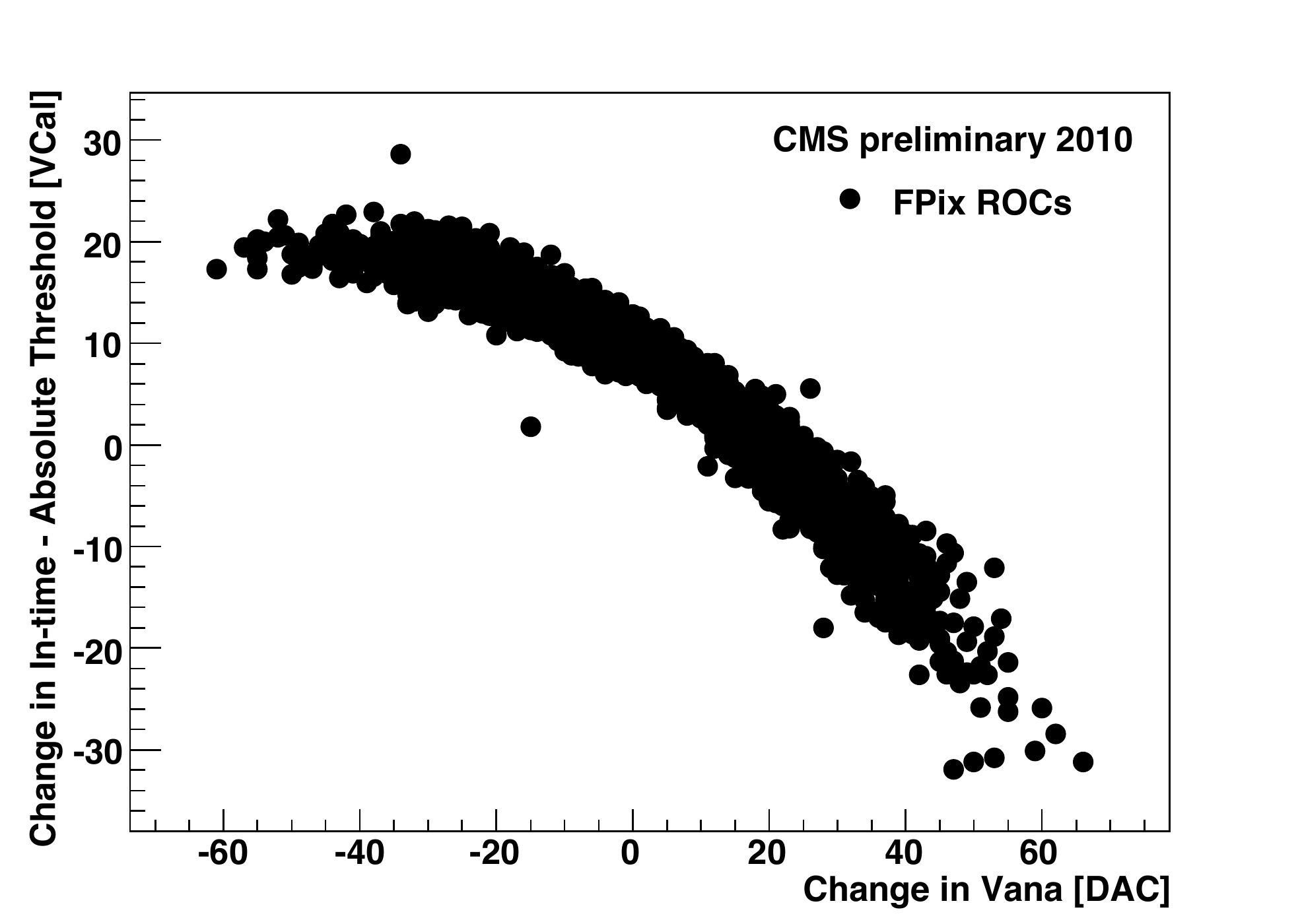}}
  \caption{Scatter plot of the change in the difference between the average in-time threshold and the approximate absolute threshold of the pixels on a ROC versus the change in Vana applied in the FPix Vana calibration.  As Vana was increased (decreased), the difference between the in-time and absolute thresholds decreased (increased) due to the change in time-walk.}
  \label{fig:plotScatter}
  \end{center}
\end{figure}

\begin{figure}[ht] 
  \begin{center}
	  \resizebox{7.5cm}{!}{\includegraphics{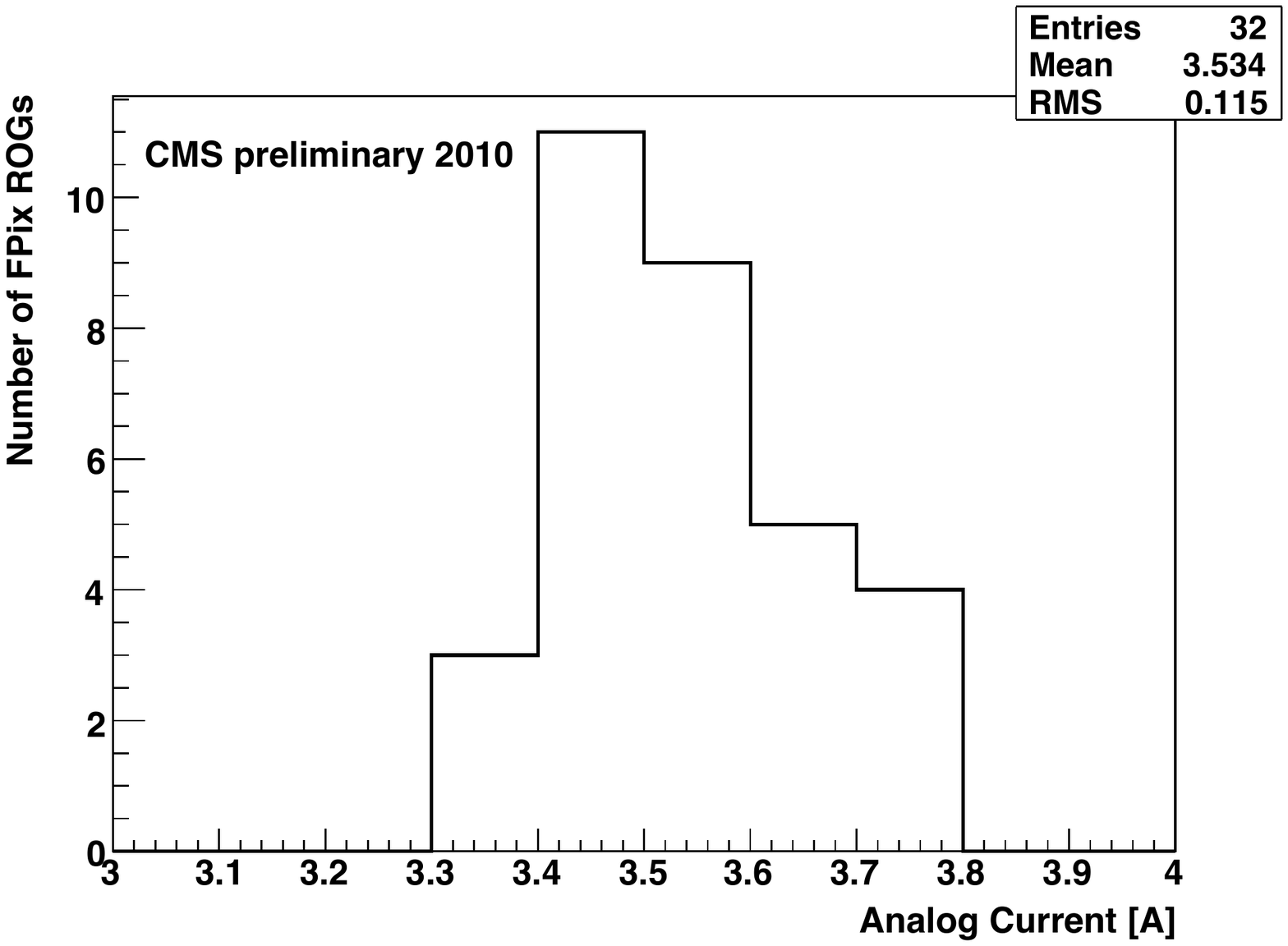}}
  \caption{Histogram of the analog currents drawn by the 32 FPix readout groups (ROGs).  Each readout group contains 135 ROCs, which makes the average analog current drawn per ROC about 26 mA.}
  \label{fig:VanaPower}
  \end{center}
\end{figure}

The Vana setting for each FPix ROC was determined without a direct measurement of the ROC's analog current.  On these ROCs, Vana was set so that the difference between the average in-time threshold and approximate absolute threshold was 12 VCal (786 electrons).  This was done in an iterative procedure where the next Vana was chosen based on the current difference.  The impact of Vana on this quantity, which quantifies the time-walk, is shown in Fig. \ref{fig:plotScatter}.  In each iteration, the absolute threshold and charge injection timing were recalibrated because they also depend on Vana.  A difference of 12 VCal was chosen because it was found to make the average analog current drawn per ROC near the FPix target of 25 mA.  Fig. \ref{fig:VanaPower} shows the resulting analog currents.  

As shown in this issue \cite{urs}, various off-line methods using collision data show that the in-time thresholds are $700$-$1000$ electrons higher than the absolute thresholds.  This is consistent with expectations from charge injection.  

\subsection{Lowering the Absolute Thresholds}
\label{subsec:33}

Increasing the cluster size from one to two pixels, so that signal interpolation can be used, increases the hit position resolution of the detector.  The cluster size was maximized by minimizing the time-walk (as described in Section \ref{subsec:32}) and then lowering the absolute thresholds.

The absolute thresholds were lowered using a ROC-based approach that works well after pixel trimming has been performed.  The absolute thresholds on a ROC should not be set below the level of cross-talk on the ROC.  When the absolute thresholds are set below this level, the ROC is overwhelmed by spurious hits that fill the double-column buffers and prevent real hits from being read out.  The absolute thresholds were set just above this failure point.

To begin the absolute threshold calibration, the absolute thresholds were set to modest values so that none of the ROCs were failing.  This can be done quickly for most ROCs by adjusting the VcThr setting on each ROC until a representative subset of its pixels are $100\%$ efficient for an injected charge of $\sim50$ VCal, for instance.  

Next, the absolute thresholds were lowered from the working point found in the first step by raising the VcThr setting on all ROCs in five steps of two DAC units.  At each step, the ROC was checked for failure by looking for failed S-Curve fits, which is one artifact of the absolute thresholds being too low.  S-Curves from a small but representative subset of the pixels on each ROC ($\sim2\%$ distributed across the ROC) were considered to save time.  The majority of ROCs failed within five steps.  

The VcThr setting of each failing ROC was then set to four DAC units below its failing point, and the VcThr setting of each ROC that never failed was set to two DAC units below the highest setting tested.  After moving the ROCs to their working settings all at once, a small number failed and were manually tuned.  Then, in an independent test known as PixelAlive, the hit efficiency for charge well over threshold was measured for every pixel.  Any ROC showing inefficiencies due to the absolute thresholds being too low, such as inefficient double-columns resulting from filled buffers, was manually tuned.  

\begin{figure}[ht] 
  \begin{center}
	  \resizebox{7.5cm}{!}{\includegraphics{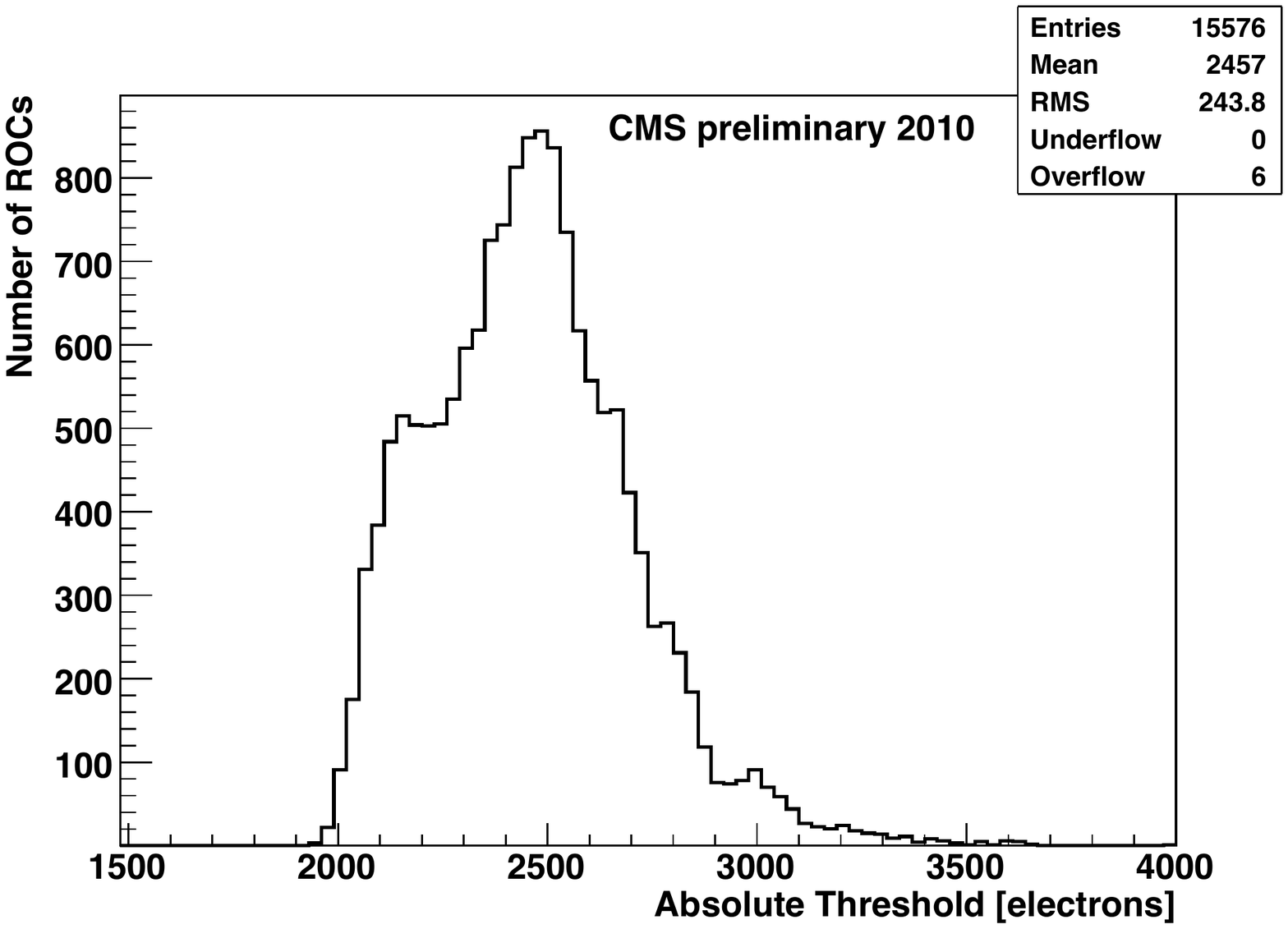}}
  \caption{Absolute threshold distribution.  Each entry is the mean absolute threshold of one ROC, which is computed from $\sim2\%$ of the pixels on the ROC.}
  \label{fig:thr}
  \end{center}
\end{figure}

The final absolute threshold distribution is shown in Fig. \ref{fig:thr}.  The mean absolute threshold is 2457 electrons, which is approximately $10\%$ of the charge collected from a minimum ionizing particle that has passed through a sensor with a vanishing incidence angle.  

This threshold was achieved without introducing a significant number of inefficient or noisy pixels.  The final number of bad pixels on ROCs included in the readout was measured in several ways.  Noisy pixels were identified and masked during cosmic ray data taking.    Pixels inefficient to charge injection were identified using the PixelAlive test; this test does not identify dead sensors or poor connections between the sensor and ROC.  Dead pixels were identified by their lack of hits in high-statistics collision data.  The final numbers are shown in Table \ref{tab:dead}.

\begin{table}[ht]
  \begin{center}
        \caption{Total number of bad pixels on ROCs included in the readout according to several tests.  $98.1\%$ of all ROCs are included in the readout. }
        \label{tab:dead}
        \begin{tabular}{l c c}
        \hline
                                       &BPix& FPix\\
        \hline
                Noisy in cosmic data and masked         & 616            &30\\
                Inefficient to charge injection           & $\sim3$k            &$\sim3$k \\
               Dead in collision data            &$\sim7.5$k             &$\sim4$k \\
        \hline
        \end{tabular}
  \end{center}
\end{table}

\subsection{Noise}
\label{sec:34}

\begin{figure}[ht] 
  \begin{center}
	  \resizebox{7.5cm}{!}{\includegraphics{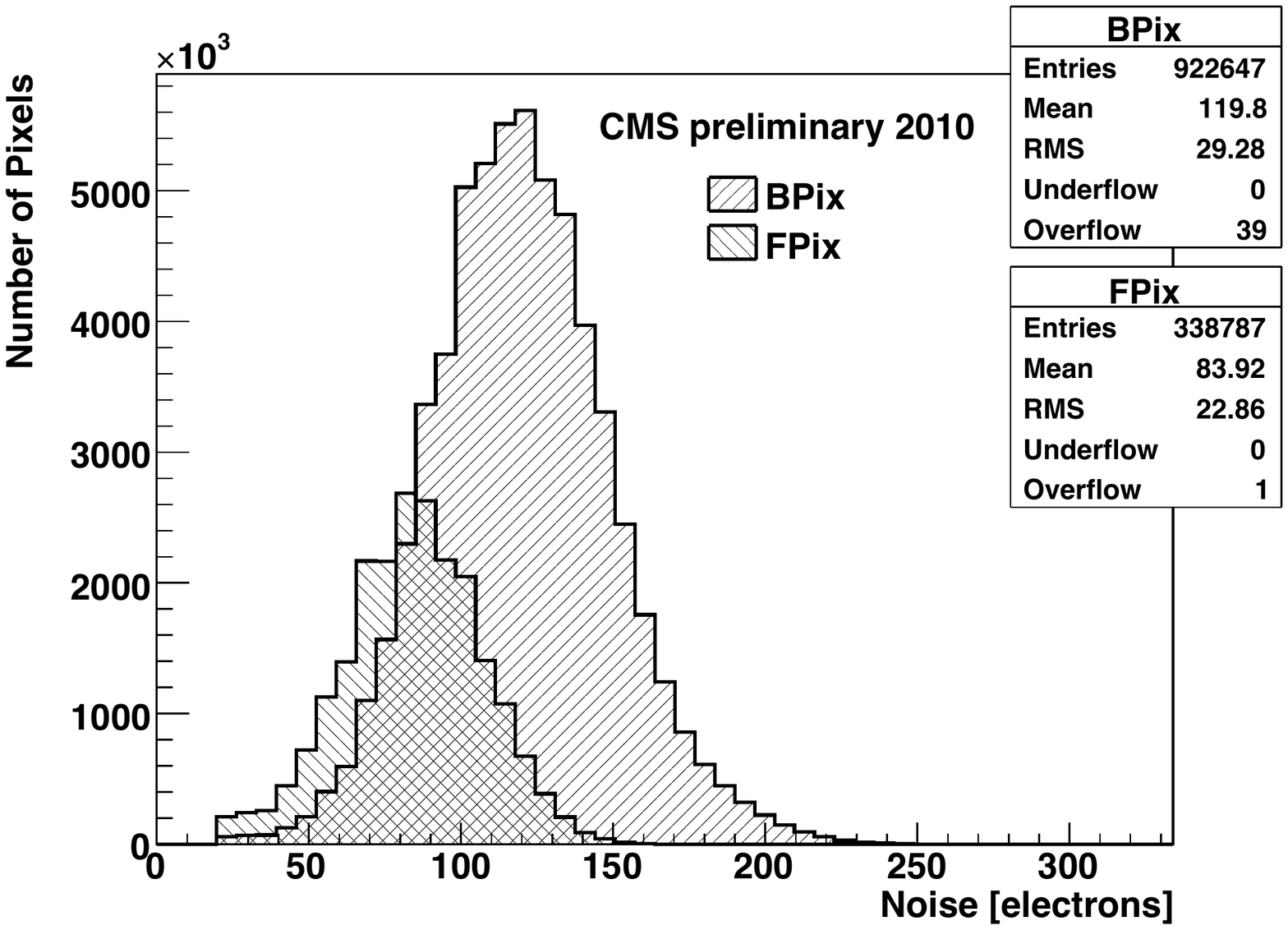}}
  \caption{BPix and FPix noise distributions obtained from $\sim2\%$ of the pixels on each ROC.}
  \label{fig:Noise}
  \end{center}
\end{figure}

The noise of a pixel is equal to the width of the turn-on region of its S-Curve, which is taken as two times the standard deviation of the Gaussian function that would result from differentiating the error function fit.  The BPix and FPix noise distributions are shown in Fig. \ref{fig:Noise}.  The BPix mean noise is 120 electrons, and the FPix mean noise is 84 electrons.   The noise of each pixel is well below the absolute threshold set just above the level of cross-talk, so it does not negatively impact the performance of the detector.

\section{Analog Response Calibrations}
\label{sec:4}

The analog response of each ROC was optimized by maximizing the linearity and range of the gain.  The gain of a pixel is defined as the pulse height versus injected charge.  It is important for the gain to be as linear as possible because it is parameterized with a linear function in the off-line reconstruction to limit the number of parameters it must store. 

The linearity of the gain was maximized through the calibration of two 8-bit DACs known as VHldDel and Vsf.  VHldDel determines a delay applied to each pulse before its height is sampled and held in a capacitor until the double-column is read out.  VHldDel was set so that the maximum of a pulse is sampled.  Vsf regulates the voltage applied to the sample and hold circuit.  Vsf was calibrated differently in the BPix and FPix.  The BPix algorithm is based on the observation that linearity increases as Vsf is increased.  Vsf was increased on each BPix ROC until the average linearity reached a target value or a current limit was reached.  In the FPix, Vsf was set so that the pulse height when VHldDel is set to its minimum is equal to the pulse height when VHldDel is set to its maximum; this was also observed to produce good linearity.  

The range of the gain of one pixel is defined as the difference between the pulse height due to the maximum injected charge and the pulse height due to injected charge just above threshold.  The range depends on several DACs, but it can be maximized by calibrating only two of them on a ROC-by-ROC basis after the rest have been set to compatible values.  VIbias\_PH and VoffsetOp, both 8-bit DACs, were calibrated.  VIbias\_PH applies a gain to the pulse height, and VoffsetOp applies an offset.  The ranges of a small but representative subset of the pixels on each ROC were measured in a two-dimensional scan of these DACs, and the settings that produced the largest ranges within the range of the FEDs' ADCs were chosen.  

The final gain parameters resulting from these calibrations are presented in this issue \cite{urs}.

\section{Regular Recalibrations}
\label{sec:5}

All of the ROC DAC settings and most of the settings in the pixel data acquisition system will not need to be recalibrated until the detector's operating temperature is changed or significant radiation damage is accumulated.  This is not foreseen to occur before 2012.

There are several FED parameters that are recalibrated on a regular basis to account for environmental changes and to monitor the detector's status.  The most frequently changed parameters are offsets in the optical receiver of each FED channel.  These are recalibrated approximately once per week when temperature changes at the laser drivers near the front end shift the signal beyond what can be handled by an automatic correction in the FED.  The automatic correction can account for temperature changes of $2-3$ $^{\circ}$C.  

Approximately two to four times per month, the parameters necessary to decode the addresses of hit pixels are remeasured.  The address parameters are relatively stable, so they are often remeasured only to check for problems.  Finally, approximately once per month, the optimal phase of each FED channel's ADC is remeasured.  The phase parameters are stable, and therefore, this calibration is performed mostly as a check.

For more detail on these calibrations, see \cite{danekComm}.

\section{Online Experience with Beam Background}
\label{sec:6}
Showers of particles that graze the detector along the beam axis and give rise to occupancies much larger than those expected from collisions at the LHC's full design luminosity have been observed.  These beam background events are consistent with expected interactions between beam and gas in the LHC beam pipe.

On the order of $10$k$-100$k pixels can register a hit in just one beam background event.  When they are coincident with a trigger, all of the hits must be read in by the corresponding FED channels, which can take up to tens of milliseconds.  This imposes a challenge to maintaining event synchronization because the events that follow come long after the affected FED channels expect them.  A two-part solution has been implemented to maintain synchronization when this occurs.  First, a mechanism to ignore the delayed events, rather than confuse them with expected events, was implemented.  The number of events that must be ignored increases with the trigger rate.  The second part of the solution was implemented to prevent this number from climbing too high.  When the number of events that must be ignored is greater than a configurable number, $N$, CMS triggers are stopped until the affected FED channels can regain synchronization by ignoring the next $N$ events.

\section{Conclusion}
\label{sec:conclusion}
The performance of the detector has been optimized for the first LHC run.  Calibrations performed in 2009 improved the threshold and analog response of the detector.  A few FED parameters are recalibrated on a regular basis to account for environmental changes and to monitor the detector's status.  A mechanism to maintain event synchronization in the FED while it reads in large beam background events has been implemented.

\section*{Acknowledgments}
\label{sec:acknowledgments}
The author would like to thank Anders Ryd, Danek Kotli\'nski, Will Johns, and Gino Bolla for helpful discussions. This work was supported by the U.S. National Science Foundation under NSF PHY-0846388 and NSF PHY-0757894 and by the U.S. Department of Energy under contract No. DE-AC02-07CH11359. 




%



\end{document}